\documentclass[preprint,showpacs,preprintnumbers,amsmath,amssymb,floatfix]{revtex4}
\usepackage{graphicx}% Include figure files
\usepackage{dcolumn}% eqnarray table columns on decimal point
\usepackage{bm}% bold math
\usepackage{verbatim}

\def\hf{{\frac{1}{2}}}

\newtheorem{thm}{Theorem}[section]
\newtheorem{cor}[thm]{Corollary}
\newtheorem{lem}[thm]{Lemma}
\newtheorem{defn}[thm]{Definition}

\begin{document}

\bibliographystyle{plainnat}
\title{Negative specific heat for quasi-2D vortex structures in electron plasmas: an explicit, closed-form derivation}
\author{T. D. Andersen}
\email{andert@alum.rpi.edu}
\author{C. C. Lim}
\email{limc@rpi.edu}
\affiliation{Mathematical Sciences, RPI, 110 8th St., Troy, NY, 12180}
\date{\today}

\begin{abstract}
Negative specific heat is a dramatic phenomenon where processes decrease in temperature
when adding energy.  It has been observed in gravo-thermal collapse of
globular clusters.  We now report finding this phenomenon in bundles of nearly parallel,
 periodic, single-sign generalized vortex filaments in the electron magnetohydrodynamic (EMH) model for the unbounded plane under strong magnetic
 confinement.  We derive the specific heat using a steepest descent method and a mean field property. Our derivations show that as temperature increases, the overall size of the system increases
 exponentially and the energy drops.  The implication of negative specific heat is a runaway
 reaction, resulting in a collapsing inner core surrounded by an expanding halo of filaments.
\end{abstract}
\maketitle
\section{Introduction}
While \cite{Schrodinger:1952} has proven that systems that are not isolated from the environment must have positive specific heat, 
the specific heat in isolated systems can be negative \cite{Lynden:1977}.  Negative specific heat is an unusual phenomenon first discovered in 1968 in microcanonical (isolated system) statistical equilibrium models of
gravo-thermal collapse in globular clusters \cite{Lynden:1968}.  In gravo-thermal collapse, a disordered system of stars in isolation under-goes a process of core collapse with the following steps: (1) faster stars are lost to an outer halo where they slow down, (2) the loss of potential (gravitational) energy causes the core of stars to collapse inward some small amount, (3) the resulting collapse causes the stars in the core to speed up.  If one considered the ``temperature'' of the cluster to be the average speed of the stars, this process has negative specific heat because a loss of energy results in an increase in overall temperature.

In the intervening four decades, negative specific heat has been
observed in few other places.  In a
magnetic fusion system or other thermally isolated plasma, should negative specific heat exist, the related runaway collapse could have profound implications for fusion where extreme confinement is critical to a sustained reaction.  

Our results have general applicability to vortex systems. However, in this paper, we address a plasma model known as the electron magnetohydrodynamical (EMH) model, where we report finding negative specific heat.  Our findings are based on a mean-field approach to the statistical equilibrium of the system.

Typically, magnetohydrodynamic plasma models are two fluid models, requiring equations governing the electron motion and equations governing the ion motion coupled together \cite{Uby:1995}.  The EMH model bypasses the two-fluid model by representing the electron fluid and the magnetic field as a single, generalized fluid with a neutralizing ion background that is stationary on the timescale chosen.  

The EMH model takes the magnetic field, ${\bf B} = \nabla\times {\bf A}$, and the charged fluid vorticity, ${\bf \omega} = \nabla\times{\bf v}$, and combines them into a general vorticity field $\Omega = \nabla\times {\bf p}$ where the generalized momentum, ${\bf p} = m{\bf v} - e{\bf A}$, $m$ is the electron mass, $-e$ is the electron charge, ${\bf v}$ is the fluid velocity field, and ${\bf A}$ is the magnetic vector potential field.  For a brief overview of the model, see \cite{Uby:1995}.  A detailed model discussion can be found in \cite{Gordeev:1994}.

Our goal is to find an {\em explicit, closed-form} formula for the specific heat of this vortex model in statistical equilibrium given an appropriate definition for energy and a microcanonical (isolated) probability distribution.  Our approach is to describe the statistical behavior of a large number of discrete, interacting
vortex structures and consider the limiting case.  We hypothesize that the specific heat is negative.

\section{Quasi-2D Vortex Model in Statistical Equilibrium}
In this section we define nearly parallel vortex filaments which are the ``particles'' of our model.  They are asymptotically thin vortex tubes that are polarized to be nearly parallel to the axis of rotation (or magnetic confinement; in the case of the EMH model, both can be considered to be present).  We go on to give (but not derive) the equations of motion and Hamiltonian for these filaments and the statistical distribution for the filaments in a microcanonical distribution.

First we define nearly parallel vortex filaments:
\begin{defn}
Nearly parallel vortex filaments are smooth curves with a complex parameterization $\psi_i(\tau,t)$ where $\psi_i(\tau,t)=x_i(\tau,t) + iy_i(\tau,t)$ and $\tau\in[0,1]$, and $t$ is time.  They are periodic, $\psi_i(0,t)=\psi_i(1,t)$.  

If we take any two values of $\tau$, $\tau_0$ and $\tau_1$ such that $\tau_0<\tau_1$, and let
$\Delta\tau=\tau_1-\tau_0$ such that $\Delta\tau\in O(\epsilon)$ where $\epsilon\ll 1$, then
for any filament $i$, the amplitude is of order $\epsilon^2$, i.e. $|\psi_i(\tau_1) - \psi_i(\tau_0)|\in O(\epsilon^2)$.
\label{def:nearlyparallel}
\end{defn}  In words this means that,
for a small rise of length $\epsilon$ in the filament, the amplitude must be on the order of $\epsilon^2$.  This assumption guarantees a certain degree of straightness in the filament that allows
for the derivation of the quasi-2D equations of motion.

The $N$ coupled, non-linear Schr\"odinger equations for the motion of these curves are the following (\cite{Kinney:1993},\cite{Uby:1995},\cite{Klein:1995},\cite{Lions:2000}):
\begin{equation}
-i\frac{\partial\psi_i}{\partial t} = \frac{\partial^2\psi_i}{\partial\tau^2}+2\frac{\psi_i-\psi_j}{|\psi_i-\psi_j|^2}.
\end{equation}

The PDE leads to a convenient energy functional,
\begin{equation}
E_N = \alpha\int_0^1 \sum_{i=1}^{N} \frac{1}{2}\left|\frac{\partial \psi_i(\tau)}{\partial\tau}
\right|^2 d\tau - \frac{1}{2}\int_0^1\sum_{i=1}^N\sum_{j=1}^N \log|\psi_i(\tau) - \psi_j(\tau)|d\tau,
\label{eqn:Ham}
\end{equation} where $\alpha$ is the core-structure constant in units of energy/length.  The first term in the energy derives from a local-induction approximation (LIA) that causes Brownian variations along the length of filament in the plane.  The second term is the coupling term and results in repulsion between filaments, a typical 2D Coulomb interaction that happens only within each plane and not between planes.

The energy is the first conserved quantity.  The second conserved quantity is angular momentum,
\begin{equation}
M_N = \sum_{i=1}^N\int_0^1|\psi_i|^2 d\tau.
\end{equation}  We define the {\bf enthalpy} to be $H_N = E_N + pM_N$, where $p$ is a fixed parameter that we call {\bf pressure}.  Because energy and angular momentum are conserved, enthalpy is conserved.

We define the micro-canonical probability for the statistical equilibrium of the system of $N$ curves.  First
\begin{defn}
Let the set of states $U$ be the space of all sets of $N$ smooth, complex functions on the interval $[0,1]$, $u=\{\psi_i\}_{i=1\dots N}$, such that $\int |\psi_i(\tau)|^2 d\tau < \infty$.
\end{defn}

The functions in $U$ include both nearly parallel and not-nearly parallel functions.  However, we will argue that not-nearly parallel functions have negligible effect on the statistics.

\begin{defn}
Let $G_N : U \rightarrow [0,\infty)$ be a Gibbs density functional and $P_N : U \rightarrow [0,1]$ be a Gibbs probability density defined by the equations
\begin{eqnarray}
G_N(u) = \delta(NH_0 - H_N)\delta(NR^2 - M_N) \\
\label{eqn:gibbs}
P_N(u) = \frac{G_N(u)}{Z_N}
\label{eqn:gibbsprob}
\end{eqnarray} where $Z_N=\int G_N(u) du$ is a normalizing factor called the partition function, and $\delta$ is a Dirac-delta function.
\end{defn}
The constants $H_0$ and $R^2$ are the enthalpy and angular momentum per vortex filament per period respectively.

The density and probability density functionals define the intersection of the function-space ``areas'' of the enthalpy-surface and the angular-momentum-surface.  If the enthalpy surface is the level set $H_{surf} = \{ u | H_N = NH_0\}$ and the angular momentum surface is $M_{surf} = \{ u | NR^2 = M_N \}$, then the intersection of the two sets is $A = H_{surf} \bigcap M_{surf}$, and $Z_N$ is the size of the set $A$.  Any probability $p = \int_B P_N(u) du$ describes the size of the intersection of $B$ and $A$ normalized by the size of $A$.  The probability density describes the statistical equilibrium of the model.  

Note: An equivalent distribution, the intersection of the {\em energy} surface and the angular momentum surface, is less appropriate in this case because it does not make the pressure constant, $p$, explicit.  Also note that the angular momentum $R^2$ is an output parameter not an input.  Again, we include it to make it explicit.

\section{The Mean-field Approximation}
To find an explicit, closed-form formula for the specific heat, we need to have an explicit, closed-form formula for the partition function $Z_N$.  Current mathematical understanding makes a direct integration impossible, but an approximation can remove the difficulty.  This approximation is called a {\bf mean-field} approximation because it averages the effect of all the filaments on one and replaces the other filaments with an external field based on the average.  Because the interaction term is based on the distance between each of the filaments, we replace, at some unknown cost, this distance with an average value.  In the next section, Section \ref{sec:uniformity}, we discuss the cost of this assumption.

\begin{thm}
Assuming that for any given filaments $i$ and $j$ and plane $\tau$ the random variables $\psi_i(\tau)$ and $\psi_j(\tau)$ are uniformly distributed on a circle of radius $2R$, then
\begin{equation}
E\left[\frac{1}{4}\sum_{i=1}^N\sum_{j=1}^N\int_0^1 \log|\psi_i-\psi_j|^2 d\tau\right] = \frac{N^2}{4}\log R^2,
\end{equation} where $E$ denotes expectation value.
\end{thm}

The expectation value can be brought into the sums and integrals with no problem,
\begin{equation}
E\left[\frac{1}{4}\sum_{i=1}^N\sum_{j=1}^N\int_0^1 \log|\psi_i-\psi_j|^2 d\tau\right] =
\frac{1}{4}\sum_{i=1}^N\sum_{j=1}^N\int_0^1 E\left[\log|\psi_i-\psi_j|^2\right] d\tau.
\end{equation}

Now let $\psi_i(\tau)=z_1$ and $\psi_j(\tau)=z_2$, and consider the expectation in polar coordinates,
\begin{eqnarray}
E\left[\log|z_1-z_2|^2\right] = \frac{1}{(\pi 4R^2)^2}\int_0^{2\pi}\int_0^{2\pi}\int_0^{2R}\int_0^{2R} \nonumber\\\log\left[r_1^2 + r_2^2 - 2cos(\theta_2 - \theta_1)\right] r_1dr_1r_2dr_2d\theta_1d\theta_2.
\end{eqnarray}

This integral has been shown to evaluate to $\log R^2$ plus a constant that we can drop without loss of generality \cite{Assad:2005}.

This concludes the mean-field approximation.  The energy functional now reads:
\begin{equation}
E_N' = \left[\alpha\int_0^1 \sum_{i=1}^{N} \frac{1}{2}\left|\frac{\partial \psi_i(\tau)}{\partial\tau}
\right|^2 d\tau\right] - \frac{N^2}{4} \log R^2.
\label{eqn:HamMF}
\end{equation}  Each of our previously defined functionals, $H_N$, $G_N$, $P_N$, and $Z_N$, now has a mean-field version, and, for the sake of simplicity of notation, we drop primes and refer only to the mean-field functionals.

\subsection{A circle of radius $2R$}
\label{sec:uniformity}
The assumption of uniformity is not as drastic as it may appear.  First of all,  just because we assume a uniform distribution for the interaction energy does not make the resulting distribution uniform.  In fact, the distribution is only uniform if the self-energy is zero.  Therefore, the assumption is not that $P_N$ is uniform but that assuming that it is uniform for the sake of gaining a simpler interaction term does not change the physics significantly.  There is significant justification for this assumption.

We argue using results from related statistics.  In a previous paper \cite{Andersen:2007e}, using Monte Carlo simulations, we showed this same mean-field approximation to be extremely effective in the {\bf canonical} (non-isolated) case of nearly parallel vortex filaments with distribution $P_N^{c} = \exp(-\beta H_N)/Z_N^{c}$, where $Z_N^{c} = \sum_s \exp(-\beta H_N)$ and $\beta$ is the {\bf inverse temperature} parameter.  The canonical distribution and the micro-canonical distribution are often equivalent, and, even when they are not, their distributions can have many of the same properties.  As we will show, with the mean-field approximation in place, the formula we obtain for $R^2$ in terms of the parameters from the micro-canonical distribution, $P_N$, is identical to the formula we obtained in our previous paper for the canonical distribution, $P_N^{c}$.

NB: If the distribution is uniform with radius $2R$, then the mean angular momentum {\em must} be $NR^2$, and, if the distribution is uniform and the angular momentum is $NR^2$, then the radius of the distribution {\em must} be $2R$.  This is from the definition of $M_N$.

\section{Maximal Entropy}
The specific heat formula that we are about to derive is the specific heat of the {\bf most-probable macrostate}.  A macrostate in our case is specified by the number of particles, $N$, the enthalpy, $H_0$, and the pressure, $p$ and is a collection of microstates: $A(N,H_0,p)\subset U$.  There are many macrostates for each fixed set of parameters, and each has a different total entropy, $S$, temperature, $T$, angular momentum, $R^2$, specific heat, $c_p$, etc.  However, only one of these macrostates is likely to occur, the one with the largest entropy.  To see why note the formula for {\bf entropy-density} by Shannon \cite{Shannon:1948},
\begin{equation}
S_N(u)=-P_N(u)\log P_N(u),
\label{eqn:shannon}
\end{equation} where $u\in U$ is a microstate.  The bigger the probability, the bigger the entropy.  Therefore, the macrostate with maximum entropy is also the most-probable:
\begin{defn}
The most-probable macrostate is a macrostate $U_{mp}\subset U$ such that the entropy of $U_{mp}$ is maximal.  The {\bf maximal entropy} is defined as $S^{max}_N = \int S_N(u) du$, where the integral is over $u\in U_{mp}$.
\end{defn}
This definition is extremely valuable because it gives us a way of describing state variables such as temperature and specific heat with fixed values, rather than the true, fluctuating ones:
\begin{defn}
The {\bf inverse temperature} of the system is defined to be the change in maximal entropy with respect to total enthalpy,
\begin{equation}
\beta_0 = \frac{1}{T} = \frac{\partial S^{max}_N}{\partial (NH_0)},
\end{equation} and the specific heat (at fixed pressure $p$) the change in total enthalpy with respect to temperature at maximal entropy,
\begin{equation}
c_p = \frac{\partial (NH_0)}{\partial T} = -\beta_0^2\frac{\partial (NH_0)}{\partial \beta_0}. 
\end{equation}
\label{def:spheat}
\end{defn}  Defining these variables in terms of the now fixed maximal entropy (states of greatest disorder) and fixed enthalpy makes temperature and specific heat fixed for a given set of parameters.  Redefining state variables that are not fixed in reality into those that are is the essence of statistical mechanics.

With these definitions, we can begin to calculate the formula for the maximal entropy which will lead to specific heat in the next section.  We now limit our investigation to only the most-probable macrostate:
\begin{thm}
 Assuming that all states are in the most-probable macrostate, i.e. $U_{mp}=U$, in the limit as $N\rightarrow\infty$ and with the necessary, {\bf non-extensive} scalings $\beta_0'=\beta_0 N$, $H_0' = H_0/N$, $\alpha'=\alpha/N$, and $p'=p/N$, the maximal entropy per filament is
\begin{equation}
 S = \beta_0' H_0' + \frac{\beta_0'}{4}\log(R^2) - \frac{1}{2\alpha'\beta_0' R^2} - \beta_0'p' R^2,
\end{equation} where
\begin{equation}
 R^2 = \frac{\beta_0'^2\alpha' + \sqrt{\beta_0'^4\alpha'^2 + 32\alpha'\beta_0'^2p'}}{8\alpha'\beta_0'^2p'}.
 \label{eqn:rsq}
\end{equation}
\end{thm}

We proceed to give a brief proof of the formula:

We can define the maximal entropy in terms of the partition function: $S_N^{max} = \log Z_N$ using Equation \ref{eqn:shannon} above.  We re-write it to look like this:
\begin{equation}
 e^{NS} = \int D\psi \delta(NH_0-H_N)\delta(NR^2-M_N),
 \label{eqn:eS}
\end{equation} where $NS=S_N^{max}$ and $S$ is the maximal entropy per filament, and $\int D\psi$ is a Feynman integral operator \cite{Feynman:1948}.  When we take $N$ in infinity, we will be left with $S$ rather than $S_N^{max}$.

In order to apply steepest-descent methods, we replace the Dirac-delta function with its integral representation,
\begin{equation}
 e^{NS} = \int D\psi \delta(NR^2-M_N)\int_{\beta_0-i\infty}^{\beta_0+i\infty}\frac{d\beta}{2\pi i}e^{\beta NH_0 - \beta H_N},
 \label{eqn:eS1}
\end{equation} where $\beta_0$ is defined to be the $\beta$-value at which the integrand attains its maximum value.

The integral is clearly finite because the exponential is quadratic negative definite, and the remaining delta function is only a constraint making the domain of integration smaller.  One could argue that Fubini's theorem does not apply to functional integration, but, in the case of this particular, traditional Feynman integral, it does apply, and one could easily show it using Feynman's piecewise linear segment approximation and taking the limit as the number of segments becomes large \cite{Feynman:1948}.  The order of integration can be rearranged to give,
\begin{eqnarray}
 e^{NS} = \int_{\beta_0-i\infty}^{\beta_0+i\infty}\frac{d\beta}{2\pi i}e^{\beta NH_0}\int D\psi \delta(NR^2-M_N) e^{- \beta H_N}.
 \label{eqn:eS2}
\end{eqnarray}  This rearrangement has produced an expression that is familiar from our previous papers \cite{Andersen:2007a} and \cite{Andersen:2007e}, namely the canonical partition function, 
$Z_{can} = \int D\psi \delta(NR^2-M_N) e^{-\beta H_N}$.  

In those papers, we showed that if $F=-\lim_{N\rightarrow\infty} \frac{1}{\beta N}\log Z_{can}$, given the scaling that $\beta' = \beta N$, $\alpha' = \alpha/N$, and $p'=p/N$, then
\begin{equation}
 F = p'R^2 - 1/4\log R^2 + \frac{1}{2\alpha'\beta'^2 R^2},\nonumber
\label{eqn:freeEnergy2}
\end{equation} where
\[
 R^2 = \frac{\beta'^2\alpha' + \sqrt{\beta'^4\alpha'^2 + 32\alpha'\beta'^2p'}}{8\alpha'\beta'^2p'},
 \label{eqn:rsqFE}
\]  In order to use our previous results, we need to take the limit (and scalings) on Equation \ref{eqn:eS2}:
\[
S = \lim_{N\rightarrow\infty} \frac{1}{N}\log \int_{\beta_0'-i\infty}^{\beta_0'+i\infty}\frac{d\beta'}{2\pi i}e^{\beta' NH_0'}Z_{can}.
\]  The steepest-descent argument from \cite{Berlin:1952} and \cite{Horwitz:1983} says we can replace all the instances of $\beta'$ with $\beta_0'$ and lose the integral over $\beta'$,
\begin{eqnarray}
S &=  \beta_0'H_0' + \lim_{N\rightarrow\infty} \frac{1}{N}\log Z_{can}\nonumber \\
&= \beta_0'H_0' - \beta_0' F,\nonumber
\end{eqnarray} which proves the formula.

\section{An Astronomical Anomaly: Negative Specific Heat}
Knowing the maximal entropy for any system for a given set of parameters, allows one to determine any of its state variables.  However, in our case we have derived a maximal entropy that is {\em dependent} on a state variable, temperature.  Therefore, we cannot determine the specific heat until we can define the temperature in terms of the input parameters.

We find the unknown multiplier, $\beta_0$, by relating the enthalpy per filament parameter, $H_0$, to the mean enthalpy: $NH_0 = \langle H_N \rangle$ \cite{Horwitz:1983}.
\begin{thm}
The enthalpy, $H_0'$, for the infinite-$N$ system (as defined above) is related to the inverse temperature for that system by the following:
\begin{equation}
 H_0' = \frac{\partial}{\partial\beta_0'}\left(-\frac{\beta_0'}{4}\log R^2 + \frac{1}{2\alpha'\beta_0' R^2} + \beta_0'p'R^2\right).
 \label{eqn:energyprime}
\end{equation}
\end{thm}

By definition the average enthalpy is given by
\begin{equation}
 \langle H_N\rangle = \frac{\int D\psi H_N\delta(NH_0-H_N)\delta(NR^2-M_N)}{\int D\psi \delta(NH_0-H_N)\delta(NR^2-M_N)}, 
\end{equation} which has the following integral representation:
\[
 \langle H_N \rangle = \frac{\int D\psi\int_{\beta_0-i\infty}^{\beta_0+i\infty}\frac{d\beta}{2\pi i}e^{\beta NH_0}\left(-\frac{\partial}{\partial\beta}e^{-\beta H_N}\right)\delta(NR^2-M_N) }{\int D\psi \int_{\beta_0-i\infty}^{\beta_0+i\infty}\frac{d\beta}{2\pi i}e^{\beta NH_0}e^{- \beta H_N}\delta(NR^2-M_N)},
\] where we have replaced $H_N\exp(-\beta H_N)$ with $-\partial/\partial\beta \exp(-\beta H_N)$.  Now in order to continue we need to bring the integral over $D\psi$ inside the derivative so that we can use our previous steepest-descent results to simplify the expression.  We prove that this is allowed in Appendix \ref{appA}.  

Let $H=N^{-2}H_N$.  Applying the limit and scalings as before (dropping the limits of integration), we exchange the derivative and functional integral:
\[
\langle H \rangle = \lim_{N\rightarrow\infty} \frac{1}{N}\frac{\int\frac{d\beta'}{2\pi i}e^{\beta' NH_0'}\left(-\frac{\partial}{\partial\beta'}\int D\psi e^{-\beta' H_N}\delta(NR^2-M_N)\right) }{\int\frac{d\beta'}{2\pi i}e^{\beta' NH_0'} \int D\psi e^{- \beta' H_N}\delta(NR^2-M_N)}.
\]

Again according to the steepest-descent argument in \cite{Horwitz:1983}, only values of the integrand at $\beta_0'$ have any contribution to the integral as $N$ becomes large:
\[
\langle H \rangle = \lim_{N\rightarrow\infty} \frac{1}{N}\frac{\left(-\frac{\partial}{\partial\beta_0'}\int D\psi e^{-\beta_0' H_N}\delta(NR^2-M_N)\right) }{\int D\psi e^{- \beta_0' H_N}\delta(NR^2-M_N)}.
\]

The fraction simplifies to give
\begin{eqnarray}
\langle H \rangle &= \lim_{N\rightarrow\infty} \frac{1}{N}\left(-\frac{\partial}{\partial\beta_0'}\log \int D\psi e^{-\beta_0' H_N}\delta(NR^2-M_N)\right)\nonumber\\
&= \frac{\partial}{\partial\beta_0'} \beta_0'F,
\end{eqnarray} where $F$ is defined in Equation \ref{eqn:freeEnergy2}.  (Note that direct evaluation can show that switching the limit and derivative is admissible.) Since $H=H_0'$, the formula is proven.

Based on the definition in Def. \ref{def:spheat} and the two theorems, i.e. the formula for maximal entropy and the formula for mean energy and consequently temperature, a Corollary is the formula for specific heat,
\begin{cor}
The specific heat has the form:
\begin{equation}
 c_p = \frac{\beta_0'}{4}\left(\frac{\alpha'\beta_0'^2}{\sqrt{\alpha'\beta_0'^2(\alpha'\beta_0'^2+32p')}}-1\right).
 \label{eqn:spheat}
\end{equation}
\end{cor}
A little algebra is all that is required.  Therefore, like its astronomical sibling, this vortex system has negative specific heat, a hallmark of meta-stable states (see next section for details).  We would also like to point out that the key to obtaining negative specific heat is the non-extensive nature of the system.  Extensive systems, by mathematical proof, cannot have negative specific heat \cite{Thirring:1970}.  Therefore, the scaling that we introduce to make the limit non-extensive is what breaks this proof's assumptions and allows for negative specific heat.

\section{Runaway reaction: A thought experiment}
All of these mathematics rely on the mean-field assumption, which we argued was reasonable in Section \ref{sec:uniformity}.  We cannot tell if the specific heat of the original mathematical model is negative directly, but consider that all we have done is taken a complicated expression for the interaction and reduced it to something that is intuitive, namely that as the system expands in average size the interaction increases logarithmically.  We offer the following thought experiment to demonstrate how negative specific heat causes the system to behave:  

Suppose that we have the system at an enthalpy $H_0$ and decrease the enthalpy to $H_1<H_0$.  Several possible corrections can occur in each of the following enthalpy terms: self-energy, interaction energy, and angular momentum.  Either the filaments become straighter, decreasing the self-energy, or they can move apart, decreasing interaction, or they can move closer, decreasing angular momentum.  Most likely, the correction will be a combination of the three depending on which maximizes the entropy.  There are two sources of entropy: (1) increased misalignment of the filaments (decreased straightness) (2) expansion of the system as a whole.  Some balance of the two will occur.  

Now suppose the correction proceeds as follows: the filaments closest to the origin squeeze together.  They become straighter and decrease the self-energy, increase the interaction energy, and decrease angular momentum.  The filaments further away from the origin move outwards decreasing the interaction energy but increasing the angular momentum and the entropy related to expansion.  They also increase the entropy by becoming less straight, increasing the self-energy.  The three sources of enthalpy are balanced so that the total enthalpy is reset to $H_1$ due to the decreased angular momentum and self-energy in the origin and the decreased interaction energy in the outer halo.  

The overall entropy should decrease with decreased enthalpy because we have positive temperature.  The system expanded, {\bf but} the straightness decreased at the origin, so let us say that the total entropy decreased.  (Of course, in a point vortex system, this cannot happen, but here it is possible.)  Because pressure $p$ is constant and the volume in the center decreased, the temperature there increased.  Meanwhile, the filaments that moved outward saw an increase in volume and corresponding decrease in temperature.  Because the cost of moving outward in terms of increased angular momentum is with the square of the distance, the ones moving outwards will not move outwards as much as the ones moving inwards.  Therefore, the overall temperature increases with the decreased enthalpy.  This same effect is observed in globular-clusters \cite{Lynden:1968},\cite{Lynden:1977}.  We propose that this effect will show up in a numerical simulation of the micro-canonical system for certain parameter regimes.

\section{Conclusion}
We have shown a way of simplifying the interaction of a nearly parallel vortex filament model with an intuitive mean-field approximation and calculated a formula for the specific heat at maximal entropy for the system in isolation.  The specific heat was shown to be negative indicating that the vortex system is meta-stable.  With a thought experiment we proposed what the negative specific heat would mean, i.e. that vortices would separate out into a core and halo and that the core would fall in on itself.  This core would likely result in vortex merger creating a large ``hole'' at the origin with a large vortex swirl around it and many smaller filaments surrounding it.  Although this work is entirely theoretical, we propose that its results can and will be observed in computational and experimental settings.

In the introduction we suggested the EMH model as an application because of its clear relationship to magnetic nuclear fusion in which confinement is the key to attaining a self-sustaining fusion reaction.  The runaway collapse we propose may lead to such a reaction.

\appendix
\section{Proof of Interchange of Derivative and Functional Integral}
\label{appA}
The following lemma relies on what is known as the broken segment or piecewise linear approximation of the Feynman paths (\cite{Feynman:1948},\cite{Lions:2000}).  Because functional integrals require different and less familiar mathematical machinery than ordinary iterated integrals, the broken segment model provides a way to prove things about functional integrals of our particular type (rather than general functional integrals) using ordinary and familiar integral theorems (e.g. Fubini's).  To get the broken segments, we take the complex function $\psi(\tau)$ and approximate it with a piecewise linear vector, $\Psi$, of length $M$, i.e. $\Psi = (\psi(\tau_1),\psi(\tau_2),\dots,\psi(\tau_M))$ where $\tau_{i+1} = \tau_i + 1/M$ and $\tau_{M+i} = \tau_i$.  The functional integral operator $\int D\psi = \lim_{M\rightarrow\infty} \int d\Psi/a^M$, where $a$ is a necessary scaling factor.

\begin{lem}
The following equation is true:
\begin{eqnarray}
 \lim_{M\rightarrow\infty} &\int d\Psi/a^M \frac{\partial}{\partial\beta}e^{-\beta H_N(M)}\delta(NR^2-M_N(M)) \nonumber\\&= \frac{\partial}{\partial\beta}\exp(-\beta NF),
\end{eqnarray} where $a=\pi/(\alpha\beta_0 M)$.
\end{lem}

 There are two separate issues here:
\begin{enumerate}
 \item whether the derivative can be brought out of the integral: \begin{eqnarray}
\int d\Psi/a^M &\frac{\partial}{\partial\beta}e^{-\beta H_N(M)}\delta(NR^2-M_N(M)) \nonumber\\&= \frac{\partial}{\partial\beta}\int d\Psi/a^M e^{-\beta H_N(M}\delta(NR^2-M_N(M)),
\label{eqn:lemmaEqn1}
\end{eqnarray}
\item whether the derivative can be brought out of the limit:
\begin{eqnarray}
\lim_{M\rightarrow\infty} &\frac{\partial}{\partial\beta}\int d\Psi/a^M e^{-\beta H_N(M)}\delta(NR^2-M_N(M)) \nonumber\\&= \frac{\partial}{\partial\beta}\lim_{M\rightarrow\infty} \int d\Psi/a^M e^{-\beta H_N(M)}\delta(NR^2-M_N(M)).
\label{eqn:lemmaEqn2}
\end{eqnarray}
\end{enumerate}

Equation \ref{eqn:lemmaEqn1} is true if and only if 
\begin{equation}
 \int d\Psi/a^M e^{-\beta H_N(M)}\delta(NR^2-M_N(M)) < \infty,
\end{equation} and the integrand is differentiable.
The equation is integrable because the integrand is positive definite (and positive).  Therefore, it is a simple Gaussian
integral.  The smoothness of the integrand guarantees differentiability.

For \ref{eqn:lemmaEqn2}, we prove it explicitly, i.e. calculate the derivative first and then
take the limit.  If $M$ is large, we can make the approximation:
\begin{equation}
 \int d\Psi/a^M e^{-\beta H_N(M)}\delta(NR^2-M_N(M)) \approx e^{-\beta NF(M)},
\end{equation} where
\begin{equation}
 F(M) = pR^2 - N/4\log R^2 - \frac{M^2\alpha}{\beta} R^2(\eta_0-1) - \frac{M}{\beta}\log(\eta_0 + (\eta_0^2 - 1)^\hf)
\end{equation} and
\begin{equation}
\eta_0 = \sqrt{\frac{1}{(M\alpha\beta R^2)^2} + 1}.
\end{equation}

Taking the derivative,
\begin{equation}
 -\frac{\partial}{\partial\beta} e^{-\beta NF(M)} = NF(M)e^{-\beta NF(M)} F_{\beta}(M),
\end{equation} it is trivial to show that
\begin{equation}
 \lim_{M\rightarrow\infty} NF(M)e^{-\beta NF(M)} F_{\beta}(M) = NFe^{-\beta NF} F_{\beta} = -\frac{\partial}{\partial\beta} e^{-\beta NF}.
\end{equation}

Acknowledgment:
This work is supported by ARO grant W911NF-05-1-0001 and DOE grant 
DE-FG02-04ER25616.

\bibliography{pimcposter}
\end{document}